\documentclass[12pt]{JHEP3}

\usepackage{epsfig}

\newcommand{\half}{\frac12}

\newcommand{\del}{\partial}

\newcommand{\tphi}{{\tilde\phi}}

\newcommand{\diag}{{\rm diag}}
\newcommand{\nn}{\nonumber}

\newcommand{\eref}[1]{Eq.~(\ref{#1})}

\def\be{\begin{equation}}
\def\ee{\end{equation}}
\def\bea{\begin{eqnarray}}
\def\eea{\end{eqnarray}}

\def\tr{{\bf\, tr \,}}
\def\hBPS{{${\frac12}$-BPS}}
\def\tphi{{\tilde\phi}}
\def\tOmega{{\tilde\Omega}}
\def\tS{{\tilde S}}
\font\cmss=cmss10 
\font\cmsss=cmss10 at 7pt
\def\IZ{\relax\ifmmode\mathchoice
{\hbox{\cmss Z\kern-.4em Z}}{\hbox{\cmss Z\kern-.4em Z}}
{\lower.9pt\hbox{\cmsss Z\kern-.4em Z}}
{\lower1.2pt\hbox{\cmsss Z\kern-.4em Z}}\else{\cmss Z\kern-.4em
Z}\fi}
\def\IB{\relax{\rm I\kern-.18em B}}
\def\IC{{\relax\hbox{\kern.3em{\cmss I}$\kern-.4em{\rm C}$}}}
\def\ID{\relax{\rm I\kern-.18em D}}
\def\IE{\relax{\rm I\kern-.18em E}}
\def\IF{\relax{\rm I\kern-.18em F}}
\def\II{\relax{\rm I\kern-.18em I}}
\def\Id{\relax{1\kern-.32em 1}}
\def\IG{\relax\hbox{$\inbar\kern-.3em{\rm G}$}}
\def\IR{\relax{\rm I\kern-.18em R}}

\def\fourth{\frac14}

\title{Bubbling Orientifolds}
\author{Sunil Mukhi\,\footnote{Email: \tt mukhi@tifr.res.in}\,
and Mikael Smedb\"ack\,\footnote{Email: \tt smedback@theory.tifr.res.in}\\
\it Tata Institute of Fundamental Research,\\
\it Homi Bhabha Rd, Mumbai 400 005, India}
\abstract{We investigate a class of \hBPS\ bubbling
geometries associated to orientifolds of type IIB string theory and
thereby to excited states of the $SO(N)/Sp(N)$ ${\cal N}=4$
supersymmetric Yang-Mills theory. The geometries are in correspondence
with free fermions moving in a harmonic oscillator potential on the
half-line. Branes wrapped on torsion cycles of these geometries are
identified in the fermi fluid description. Besides being of intrinsic
interest, these solutions may also occur as local geometries in flux
compactifications where orientifold planes are present to
ensure global charge cancellation. We comment on the extension of
this procedure to M-theory orientifolds.}

\preprint{hep-th/0506059\\ TIFR/TH/05-19}


\begin{document}

\section{Introduction}

The recent discovery \cite{Lin:2004nb} of an infinite set of \hBPS\ geometries
of type IIB supergravity is of great interest. This discovery provides
a correspondence between semiclassical states of the matrix harmonic
oscillator and
\hBPS\ geometries, in which the ground state of the matrix model
(equivalent to a fermi fluid in a harmonic oscillator potential)
corresponds to the well-known $AdS_5\times S^5$ geometry. Also, the
$pp$-wave geometry emerges in the limit of relativistic fermions.  A
similar construction also exists for \hBPS\ backgrounds of 11-dimensional
supergravity.

These geometries can have various interpretations, as D-branes and/or
giant gravitons/dual giant gravitons, depending on the typical sizes
of various regions. Their relation to a fermi fluid profile arises
from the fact that one can pick an arbitrary shading of the complex
plane into black and white regions (denoting occupied/unoccupied
regions of fermion phase space) and input this data to construct a
\hBPS\ supergravity solution with fluxes. The matrix oscillator in
turn arises\cite{Corley:2001zk,Berenstein:2004kk} (see also
Ref.\cite{Caldarelli:2004ig}) in the ${\cal N}=4$ supersymmetric
Yang-Mills theory on a D3-brane wrapped on $S^3$, which has a coupling
of the form $\int
\sqrt{g}R\,\phi_i^2$ for each complex scalar field $\phi_i$ of the
${\cal N}=4$ supermultiplet.  

This discovery provides a somewhat new insight into holography. Not
only do \hBPS\ operators of the ${\cal N}=4$ SYM theory correspond to
excited states of supergravity on $AdS_5\times S^5$, as was already
known for a long
time\cite{Maldacena:1997re,Witten:1998qj,Gubser:1998bc}, but they can
also be identified with entirely {\it new} \hBPS\ geometries\footnote{
The LLM framework thus provides a tool for studying supergravity
excitations of $AdS_5 \times S^5$ when the backreaction is large. In
other circumstances, when the backreaction is limited, other
techniques may be more useful in the holographic context. One such
example is the computations of anomalous dimensions in super
Yang-Mills theory using spin chains, as pioneered by
\cite{Minahan:2002ve} and reviewed for example in
Refs.\cite{Tseytlin:2003ii,Beisert:2004ry}.
}. These geometries can possess quite different topologies and one can
understand topology change as the process of fermi fluid profiles
merging and separating\cite{Lin:2004nb,Horava:2005pv}. The phase space
of the fermi fluid is realised on 2 of the 9 space dimensions of
10-dimensional type IIB supergravity. This implies that the gravity
configuration space is partly noncommutative, explanations of which
have been proposed in Refs.\cite{Mandal:2005wv,Grant:2005qc} (related
observations can be found in Ref.\cite{Dhar:2005qh}).

In the present work we would like to investigate the bubbling geometry
construction in the presence of an intrinsically stringy effect,
namely the orientifold plane. This ubiquitous object has led to
important insights in the
past\cite{Polchinski:1998rq,Johnson:2003gi}. In gauge theory terms it
can be used to modify the gauge group from $SU(N)$ to
$SO(2N),SO(2N+1)$ or $Sp(2N)$ without breaking any supersymmetry. From
the gravity point of view, parallel orientifold planes and D-branes
break the same supersymmetries, so it is not surprising that we will
find bubbling orientifolded geometries that are \hBPS. This adds a
nontrivial class of new geometries to the ones proposed in
Ref.\cite{Lin:2004nb}. The ``ground state'' geometry in this case is
$AdS_5\times RP^5$, an example studied in depth in Ref.\cite{Witten:1998xy}.

Besides providing more general examples of ``bubbling geometries'',
the introduction of orientifold planes in this context is likely to be of
practical interest. For compactifications of string theory with
space-filling branes and/or fluxes, it is well-known that the absence
of tadpoles requires space-filling orientifolds. In
Ref.\cite{Verlinde:1999fy}, it has been noted that the $T^6/Z_2\times
R^{3,1}$ orientifold compactification has precisely $AdS_5\times RP^5$
as the local geometry around the space-filling D3-branes if they are
brought near an orientifold. So besides representing a \hBPS\ geometry
by itself, $AdS_5\times RP^5$ can also be thought of as a possible
local geometry for a \hBPS\ compactification. This suggests a more
general class of \hBPS\ compactifications (involving both branes and
fluxes) for which the spacetimes discussed in the present paper could
appear as local geometries.

In what follows we first briefly review bubbling geometries as well as
orientifold 3-planes. Following this we describe the bubbling
orientifold geometries and their realisation in terms of free fermions
on a phase space that is the upper half-plane. We will see that the
orientifold plane can be interpreted as the wall which truncates a
regular harmonic oscillator potential to a ``half-oscillator'', a
familiar problem in quantum mechanics. We comment on the topology and
other characteristics of the most general bubbling solutions,
including discrete torsion and the possibility of two different types
of topology change. Finally we briefly examine the analogue solutions
in $M$-theory. In the Appendix, we collect some useful facts about
random matrices in a harmonic oscillator potential, for the cases when
the matrices are in the Lie algebra of $SO(2N+1),Sp(2N),SO(2N)$.
These matrix models compute the \hBPS\ states of ${\cal N}=4$
super-Yang-Mills theory with the corresponding gauge groups.

\section{Bubbling Geometries: A Brief Review}
\label{briefreview}

The type IIB \hBPS\ bubbling geometries of Ref.\cite{Lin:2004nb} can
be summarised in a family of classical solutions of type IIB
supergravity given in terms of a single function $z(y;x_1,x_2)$. The
metrics contain two 3-spheres $S^3$ and $\tS^3$ and are given by:
\bea
\label{tenmetric}
\nonumber
ds^2 &=& -\left[{y^2\over \fourth-z^2}\right]^\half\,(dt+V_i\,dx_i)^2 
+ \left[{\fourth-z^2\over y^2}\right]^\half\,(dy^2 + dx_i\,dx_i)\\ &&+
\left[y^2\,{\half+z\over\half-z}\right]^\half\,d\Omega_3^2 + 
\left[y^2\,{\half-z\over\half+z}\right]^\half\,d\tOmega_3^2 
\eea
Additionally the solution has suitable 5-form fluxes. The function
$z=z(y;x_1,x_2)$ satisfies:
\be
(\del_y^2 + \del_i\del_i) z -{1\over y}\,\del_y z =0
\ee
and $V_i$ is a vector field determined in terms of $z$ via:
\be
y\,\del_y V_i = \epsilon_{ij}\,\del_j z,\quad 
y\,\del_{[i} V_{j]} =\epsilon_{ij}\,\del_y z
\ee

>From the metric, it is clear that the range of the $y$ coordinate is
restricted to $y\ge 0$. Also as $y\to 0$, smoothness of the metric
requires $z\to\pm\half$ with $y/\sqrt{\half\mp z}$ fixed.  Then if
$z\to+\half$, the 3-sphere $\tS^3$ with metric proportional to
$d\tOmega_3^2$ shrinks to zero size, while the other 3-sphere $S^3$
with metric $d\Omega_3^2$ remains finite. When $z\to-\half$ the
reverse is true, and it is $S^3$ that shrinks to zero size. More
details are given in Ref.\cite{Lin:2004nb}. The geometries are
parametrised by drawing smooth contours in the $x_1,x_2$ plane which
demarcate regions of $z=-\half$ (black) from regions of $z=+\half$
(white). In turn, these regions can be interpreted as semiclassical
phase-space profiles of a fermi fluid where the black regions are
occupied and the white regions are unoccupied.  The constant
phase-space densities of these droplets map to constant densities in
the $x_1,x_2$ plane. This follows from the fact that the flux through
a sphere formed by drawing a surface ending on a black/white
region can be converted to an integral of a constant flux density over
the region.

Fermi fluid profiles consisting of a black disc with a circular
boundary correspond to the $AdS_5\times S^5$ solution. Deformations of
this by adding/removing thin concentric shells correspond to giant
gravitons or dual giant gravitons in this spacetime. But the most
generic geometry can have very little to do with $AdS_5\times S^5$ and
may contain, for example, an arbitrary number of $S^5$ factors.

The bubbling construction can be extended to \hBPS\ solutions of
11-dimensional supergravity. In this case one finds metrics
parametrised in terms of a function $D(y;x_1,x_2)$. In what follows, we denote
$\del_yD$ by $D'$. These metrics contain a five-sphere $S^5$ and a
two-sphere $\tS^2$, and are given by:
\bea
\label{elevenmetric}
\nonumber
ds^2 &=& -4\left[{y\over D'(1-yD')^2}\right]^{1\over 3}(dt + V_i\,dx^i)^2
+ \left[{{D'}^2(1-yD')\over y^2}\right]^{1\over 3}\Big(dy^2 +
e^D(dx_1^2 + dx_2^2)\Big)\\
&& + 4 \left[{y(1-yD')\over D'}\right]^{1\over 3}d\Omega_5^2
+ \left[{y^2D'\over (1-yD')}\right]^{2\over 3}d\tOmega_2^2
\eea
along with a suitable 4-form flux. The function $D(y;x_1,x_2)$
satisfies the three-dimensional Toda equation:
\be
\del_i\del_i D + \del_y^2 e^D =0
\ee
and the vector field is determined in terms of $D$ by:
\be
V_i = \half \epsilon_{ij}\del_j D
\ee
In this class of metrics too, the range of the coordinate $y$ is
restricted to $y\ge 0$ and as $y\to 0$, the function $D$ must obey one
of two boundary conditions. One possibility is $D'\sim y$ while the
other is $D'\sim y^{-1}$. More precisely we have:
\bea
\nn
\hbox{(i)}&\quad& D~\to~ y^2 + g(x_1,x_2)\\
\hbox{(ii)}&\quad& D~\to~ \log y + h(x_1,x_2)
\eea
where $g,h$ are functions of $(x_1,x_2)$ satisfying the
two-dimensional Liouville equation:
\be
\del_i\del_i\, g(x_1,x_2) + e^{g(x_1,x_2)}=0
\ee
and similarly for $h(x_1,x_2)$. 

In case (i), keeping $D$ finite in the limit, it is evident from the
metric \eref{elevenmetric} that $\tS^2$ shrinks to zero size while
$S^5$ remains finite. In case (ii) we keep $D'-y^{-1}$ finite in the
limit and find that $S^5$ shrinks with $\tS^2$ remaining finite. Thus
again the $x_1,x_2$ plane is divided into two types of regions, which
can be associated to fermi fluid droplets. There is an important
subtlety\cite{Lin:2004nb}: the density of fermions in a droplet is not
constant, unlike in the type IIB case. In fact the densities of the
droplets are
\bea\nn
\hbox{(i)}&\quad& \rho(x_1,x_2)= 2\,e^{g(x_1,x_2)}\\
\hbox{(ii)}&\quad& \rho(x_1,x_2) = 2\,e^{h(x_1,x_2)}.
\eea
However, as argued by the authors of Ref.\cite{Lin:2004nb}, two
droplets related by a conformal mapping of the plane give rise to the
same \hBPS\ solution. Therefore the topology of the bubbles in the
$(x_1,x_2)$ plane is expected to be the same as that in the fermion
phase space. In particular, they demonstrate that the M-theory
pp-wave is given by one of the two boundary conditions for $x_2<0$ and
the other one for $x_2>0$, just as one would expect from the
correspondence of this system with the relativistic limit of free
fermions.

\section{The $AdS_5\times RP^5$ Orientifold}\label{sec_witten}

In Ref.\cite{Lin:2004nb}, \hBPS\ excitations of a ground state geometry
corresponding to $AdS_5 \times S^5$ in type IIB string theory were
considered, and the full back reaction on the geometry was
determined. We will instead be interested in an $AdS_5 \times RP^5$
ground state, arising from a $Z_2$ orientifold projection of the
$S^5$, as considered by Witten\cite{Witten:1998xy}.

Introducing an orientifold plane changes the gauge group on a stack
of D-branes to $SO(2N)$, $SO(2N+1)$ or $Sp(2N)$, instead of
previously $SU(N)$.  The orientifolded theory has no fixed points on
the $S^5$, so there is no open string sector.  Being an orientifold,
traversing a non-contractible loop flips the orientation of the
world-sheet (or any embedded orientable manifold). This means that
there is also no ``winding sector''. The spectrum therefore consists
only of those $AdS_5 \times S^5$ states which are invariant under the
orientifold projection, at least for weak coupling\footnote{ On the
other hand, interactions will become more complicated, due to
contributions from non-orientable world-sheets.}.

The topology of the 2-form
fields $B_{NS\,NS}$ and $B_{RR}$ is in general non-trivial,
in the sense that
the field strength $H=dB$ belongs to some
equivalence
class of the cohomology group
$H^3(AdS_5 \times RP^5,\tilde{Z})=H^3(RP^5,\tilde{Z})=Z_2$,
where $\tilde{Z}$ denotes twisted coefficients, i.e. coefficients
flipping sign along a non-contractible loop.
To preserve supersymmetry, it is necessary to choose $H=0$.
We have already said that traversing a non-contractible loop
flips the orientation of the world-sheet $\Sigma$. Again using twisted coefficients,
this means that the homology is
$\Sigma \in H_2(RP^5,\tilde{Z})=Z_2$, and we may represent the non-trivial
element by $RP^2$.
Consequently, for non-trivial choice of {\it discrete torsion}
for the field strength, the contribution to the path integral is
$e^{i \int_{\Sigma} B_{NS\,NS}} = (-1)^s$,
where $s$ is the number of $RP^2$'s making up the world-sheet.
Trivial choice of torsion always contributes a factor of unity.

There will be four distinct combinations of discrete torsion which can
be labelled by
\bea
\nn
\theta_{NS\,NS}&=& {1\over 2\pi}\int_{\Sigma} B_{NS\,NS}\\
\theta_{RR}&=& {1\over 2\pi}\int_{\Sigma} B_{RR},
\eea
so that $\theta_{NS\,NS}$ and $\theta_{RR}$ can independently take the
values $0,\half$. The $(0,0)$ theory can be shown to be $SL(2,Z)$
invariant. The Montonen-Olive duality of such a theory identifies it
as $SO(2N)$.  Turning on $\theta_{NS\,NS}=\half$ will change the gauge
group to $Sp(2N)$, which can be seen in the following way.  Feynman
diagrams of $Sp(2N)$ are obtained from those of $SO(2N)$ by sending $N
\rightarrow -N$. Since the contribution of a generic diagram goes as
$N^{2-2g-s}$, where $g$ is the genus and $s$ is the number of glued
copies of $RP^2$, we conclude that each $RP^2$ contributes a factor
$-1$. By definition of the discrete torsion, this is precisely the
effect of turning on $\theta_{NS\,NS}$. This argument therefore
indicates that turning on $\theta_{NS\,NS}=\half$ (with
$\theta_{RR}=0$ or $\half$) is equivalent to changing the gauge group
from $SO(2N)$ (or $SO(2N+1)$) to $Sp(2N)$.  The only combination of
discrete torsions which is left is
$(\theta_{NS\,NS},\theta_{RR})=(0,\half)$, which can be identified
with the ``remaining'' gauge group, $SO(2N+1)$.

Interestingly, the torsion need not be constant all over the manifold.
To explain this, some background on 3-branes in this geometry is
required.  In addition to the ordinary $D3$ brane, the $AdS_5 \times
RP^5$ theory also contains 3-branes resulting from wrapping a 5-brane
on an $RP^2$ submanifold of the $RP^5$.  Any 3-brane is localised in
one spatial direction on the $AdS_5$ space.  The $D3$ brane is a
source of 5-form flux, so the brane will act as a {\it domain wall},
separating $SU(N)$ and $SU(N+1)$ gauge theories on the boundary. On
the orientifold, the flux $N$ through $S^5$, which covers the $RP^2$
twice, will instead shift by two units.  Therefore, the gauge group
will shift between $SO(2N)$ and $SO(2N+2)$ or between $Sp(2N)$ and
$Sp(2N+2)$.  But crossing a 3-brane made by wrapping a D5 or NS5-brane
on $RP^2$ also makes the $\theta$ angle jump. This is because the
wrapped 5-brane also acts as a source for the field $H=dB$. A wrapped
$D5$ brane sources RR-form flux, and so makes $\theta_{RR}$
jump. Similarly, a wrapped $NS5$ brane has NS-flux on it, making
$\theta_{NS\,NS}$ jump upon crossing it.  We will discuss how these
features relate to the LLM description in section \ref{sec_torsion}.

In general, branes characterized by untwisted or twisted
charges can only be wrapped on non-trivial cycles with
untwisted or twisted coefficients, respectively.
In addition, topological considerations may restrict
the allowed values of the discrete torsions.
In spite of these restrictions, the
$RP^5$ geometry still introduces new types of objects,
as compared to the unorientifolded theory, such as
{\it fat strings} and {\it Pfaffian particles} (on the gauge
theory side). They can be constructed by wrapping branes on non-trivial
cycles of the $RP^5$ which lack any counterpart in the $S^5$ theory.
We refer to Ref.\cite{Witten:1998xy} for details.

Here, we will content ourselves by illustrating some of the
general features involved by discussing some properties of
the {\it baryon vertex} of $SU(N)$ gauge theory,
which does exist both before and after
the projection to the $RP^5$ geometry.
In the unprojected
case, the only non-trivial cycle is the full $S^5$ itself.
The baryon vertex is constructed by wrapping a $D5$ brane
on this $S^5$.
Strings connect it to $N$ external quarks.
Quarks are particles in the
fundamental representation of the gauge group.
The combination is fermionic, since it is made gauge invariant
by contracting the colour wave functions using
an antisymmetric tensor of order $N$.

That this makes sense from the gravity point of view can be seen
as follows. There is a coupling
$\frac{1}{2\pi} \int_{S^5 \times R} a \wedge F_5$
on the world-volume of the $D5$ brane.
As always, $\int_{S^5} F_5 = 2 \pi N$,
so the charge corresponding to the $U(1)$ gauge field $a$
gets a contribution of $N$ units due to the $D5$ brane.
Similarly, each of the $N$ fundamental strings ending on the $D5$
contributes by $-1$, making the total charge vanish, as required.

The other ends of the strings connect to quarks at the
boundary, modelled by attaching their endpoints
to a large $D3$ brane whose spatial world-volume is
of the topology $S^3 \times P$, where $P$ is a point on the $S^5$.
Placing the wrapped $D5$ at the point $Q$ in $AdS_5$ and considering the
``linking'' numbers
between the manifolds $S^3 \times P$ and $Q \times S^5$, there
will be a change of sign associated with interchanging two strings,
confirming that the strings are fermionic.

In the projected theory,
topological restrictions allow the
$D5$ to wrap the $RP^5$ an {\it even} number of times only.
Hence, the baryon vertex only couples to an even number of
quarks on the orientifold.

\section{Bubbling Orientifolds and Fermi Fluids}\label{sec_fermi}

In the previous section we have seen that the spacetime $AdS_5\times
S^5$ admits an involution that reflects all the directions of the
5-sphere and converts it into the real projective space $RP^5$. The
dual gauge theory is the ${\cal N}=4$ super-Yang-Mills theory on a set
of D3-branes parallel to an orientifold 3-plane. The gauge group can
be $SO(2N)$, $SO(2N+1)$ or $Sp(2N)$ depending on the precise type of
O3-plane. In keeping with the bubbling geometry idea, we expect that
there should be an infinite set of \hBPS\ bubbling geometries that
correspond to the excited states of this gauge theory.

To find these, we need to understand how the fermi fluid profile that
corresponds to $AdS_5\times S^5$ is affected by the orientifolding. 
This profile is a circle of radius $r_0=R_{AdS}^2$ in the fermion
phase space. In the geometry this space is realised as the $x_1-x_2$
plane. Now $AdS_5\times S^5$ is parametrised as:
\be
(ds)^2 = r_0\left(-\cosh^2\rho\,dt^2 + d\rho^2 +
\sinh^2\rho\,d\Omega_3^2 + d\theta^2 + \cos^2\theta\, d\tphi^2 +
\sin^2\theta\,d\tOmega_3^2\right)
\label{sfive}
\ee 
with $\theta\in[0,\frac{\pi}2], \tphi\in[0,2\pi]$.
In terms of embedding coordinates in an $\IR^6$, one can write:
\bea
\nn
X_1 &=& R\,\cos\gamma\,\sin\alpha\,\sin\theta\\
\nn
X_2 &=& R\,\sin\gamma\,\sin\alpha\,\sin\theta\\
\nn
X_3 &=& R\,\cos\beta\,\cos\alpha\,\sin\theta\\
\nn
X_4 &=& R\,\sin\beta\,\cos\alpha\,\sin\theta\\
\nn
X_5 &=& R\,\cos\tphi\,\cos\theta\\
X_6 &=& R\,\sin\tphi\,\cos\theta
\label{embed}
\eea
with $\alpha,\theta\in[0,\frac{\pi}2]$ and $\beta,\gamma,\tphi\in[0,2\pi]$.
The embedding condition is $\sum_i X_i^2 = R^2 \equiv r_0$.

The orientifold action on this $\IR^6$ is $X_I\to -X_I$,
$I=1,2,\cdots,6$. In terms of the angular variables this amounts to
\bea
\nn
\beta &\to& \beta+\pi\\
\nn
\gamma &\to& \gamma+\pi\\
\tphi &\to& \tphi+\pi.
\eea
Going to the $x_1-x_2$ plane of the bubbling geometry, we recall that
it is described by polar coordinates $(r,\phi)$ where 
\bea
\nn\label{polardefs}
r&=&r_0\, \cosh\rho\,\cos\theta\\
\phi &=&\tphi+t,
\eea
and $\rho,\tphi$ are the coordinates appearing in Eq.(\ref{sfive})
above. Therefore, under the orientifolding operation, the $x_1-x_2$
plane undergoes the involution $\phi\to\phi+\pi$, which is the same
as the reflection $(x_1,x_2)\to (-x_1,-x_2)$.

The precise picture is a little more complicated because at $y=0$, the
full spatial geometry is not really 2-dimensional. In the regions
where $z=\pm\half$ (the white and black regions) the geometry is
5-dimensional, and consists of the $(x_1,x_2)$ plane together with one
of the 3-spheres $S^3$ or $\tS^3$, parametrised respectively by
$d\Omega_3$ or $d\tOmega_3$, while the other 3-sphere has shrunk to
zero size.  The sphere $\tS^3$ lies inside $S^5$ (and is parametrised
by the angles $\alpha,\beta,\gamma$ in Eq.(\ref{embed})). Thus it is
inverted by the orientifold action, while the other 3-sphere $S^3$
that lies in $AdS_5$ remains unaffected. Thus, at a generic point of
the $(x_1,x_2)$ plane, the orientifold involution acts by reflecting
$(x_1,x_2)$ and simultaneously inverting $\tS^3$. In the white
regions, where $z=\frac12$, the $\tS^3$ shrinks to zero size, while in
the black regions, where $z=-\frac12$, it is $S^3$ that shrinks. It
follows that in the white regions, the orientifolding operation acts
solely by inverting the $(x_1,x_2)$ plane and turning it into
$\IC/\IZ_2$. The same is true on boundaries between the black and
white regions with $z=-\half,z=+\half$ respectively (where both
$S^3,\tS^3$ shrink). In the black region, however, one has to keep in
mind that the $\tS^3$ above a given point of the $(x_1,x_2)$ plane is
identified with a reversed $\tS^3$ above the diametrically opposite
point. Finally, at the origin $x_1=x_2=0$ which is the fixed point of
$\phi\to\phi+\pi$, the $\tS^3$ gets an antipodal identification and
becomes $RP^3$.

The above discussion was based on an involution that is a symmetry of
the $AdS_5\times S^5$ background. Therefore strictly speaking it
applies only to the simple fermi fluid profile consisting of a black
disc of fixed radius centred at the origin. But now it is clear how to
define orientifolding for the most general bubbling geometry: simply
consider all fermi fluid profiles whose boundaries are well-defined on
$\IC/\IZ_2$. Such boundaries have to be invariant under a rotation by
$\pi$ in the $(x_1,x_2)$ plane. Alternatively they may be drawn on the
fundamental domain of $\IC/\IZ_2$: the upper half-plane with the
positive and negative halves of the $x_1$ axis identified with each
other. 

What is the fermi system dual to these spacetimes? For this system,
the eigenvalue phase space should be an orbifold $\IC/\IZ_2$. This
means the position $x$ of the fermion is strictly positive (recall
that the identification between coordinate space and phase space is
$(x_1,x_2)\to (p,x)$). The identification of the positive and negative
momentum axes tells us that at $x=0$ the momentum is instantaneously
reversed. This is consistent with the harmonic oscillator being
truncated to the ``half harmonic oscillator'' with an infinite
vertical wall at $x=0$, a well-known system in quantum mechanics. Thus
we have a (Hermitian) matrix-valued particle, or equivalently
fermionic eigenvalues, moving in this potential. The fermionic wave
functions are made in the usual way out of one-particle wave
functions, the latter now being the parity-odd solutions of the full
harmonic oscillator.

>From the correspondence between states of the matrix harmonic
oscillator and \hBPS\ excitations of ${\cal N}=4$ SYM, we would expect
the appropriate matrix model for an orientifolded geometry to be
related to ${\cal N}=4$ SYM with gauge group $SO(2N)$\footnote{Or
$SO(2N+1)$ or $Sp(2N)$.}. Then, following the arguments of
Refs.\cite{Corley:2001zk,Berenstein:2004kk}, one should get a usual,
not semi-infinite, harmonic oscillator, the only change being that the
Hermitian matrix (in the algebra of $SU(N)$) is replaced by an
antisymmetric matrix (in the algebra of $SO(2N)$).

This appears to be a different theory, but in fact the two
descriptions are equivalent, for the same reason that (in flat space)
an orientifold plane projects $SU(N)$ Chan-Paton factors down to the
$SO$ or $Sp$ subgroup. Consider the random matrix model for $2N\times
2N$ real antisymmetric matrices $A$ (for simplicity we choose constant
matrices) with a potential $V(A)$:
\be
Z=\int [dA] e^{-\tr V(A)}
\ee
This is invariant under the orthogonal transformations, which can be
used to reduce the matrix $A$ to the form
\bea
\nn
\Lambda &=& \pmatrix{ \lambda_1\pmatrix{0&1\cr -1 &0} & 0 & \cdots &0\cr
0 & \lambda_2\pmatrix{0&1\cr -1 &0} & \cdots &0\cr
\cdots &\cdots &\phantom{\pmatrix{0&1\cr -1 &0}}&\cdots\cr
0 & 0 & \cdots & \lambda_N\pmatrix{0&1\cr -1 &0}}\\
&=& i\sigma_2\otimes \diag(\lambda_1,\lambda_2,\ldots,\lambda_N).
\eea
Importantly, all $\lambda_i$ are real and can be chosen to be
non-negative. For example, in each block,
a negative $\lambda_i$ can be brought to a positive one by conjugating
with $\sigma_3\otimes \Id$.

The matrix measure 
\be
[dA]\equiv \prod_{P<Q}dA_{PQ}
\ee
can then be shown to reduce to
\bea\label{measure_orth}
\nn
[dA]&=& \prod_{j<k=1}^N(\lambda_j^2-\lambda_k^2)^2\prod_{i=1}^N d\lambda_i\\
&=& \prod_{j<k=1}^N(\lambda_j-\lambda_k)^2\prod_{j<k=1}^N(\lambda_j+\lambda_k)^2
\prod_{i=1}^N d\lambda_i.
\label{antisymvdm}
\eea
The nontrivial measure factor just corresponds to $\sqrt{g}$ for the
metric $g_{IJ}$ on the space of deformations, in the coordinates made
up of the $\lambda_i$ and the orthogonal transformations\footnote{The
formula above has appeared in, for example,
Ref.\cite{Bilal:1990yy}.}. 

Once we know the metric, we easily derive the kinetic term in the
Hamiltonian for this matrix model, which is just the Laplacian on the
deformation space,
\bea
\nn
H &=& \sum_i{1\over\sqrt{g}}{\del\over\del \lambda_i}
\sqrt{g}{\del\over\del \lambda_i}= {1\over 
\prod_{j<k}(\lambda_j^2-\lambda_k^2)^2}\,
\sum_i {\del\over\del \lambda_i}
\prod_{j<k}(\lambda_j^2-\lambda_k^2)^2\,
{\del\over\del \lambda_i}\\
&=& {1\over \prod_{j<k}(\lambda_j^2-\lambda_k^2)}\,\sum_i{\del^2\over
\del\lambda_i^2} \prod_{j<k}(\lambda_j^2-\lambda_k^2).
\eea
The equality of the two lines above follows from the identity
\be
\sum_i{\del^2\over \del\lambda_i^2}\left(\prod_{j<k}(\lambda_j^2-
\lambda_k^2)\right)=0.
\ee
We see that with some changes, this works out much like the case of
Hermitian matrix models. Absorbing the factor
$\prod_{j<k}(\lambda_j^2-\lambda_k^2)$ into the wave function makes it
fermionic. All this is in agreement with our argument that
orientifolding confines the coordinates of the free fermions to a
half-line. In fact it tells us something more precise -- from
Eq.(\ref{antisymvdm}), we see that the positive
``eigenvalues''\footnote{One should keep in mind that the actual
eigenvalues are $\pm i\lambda_i$.}\ $\lambda_i$ not only repel each
other, but also repel their images $-\lambda_i$. This is just what we
would expect in the presence of an orientifold plane.

In the above we have mostly focused on the ``standard'' orientifold
action that leads to the gauge group $SO(2N)$ on the gauge theory
side. The above procedure admits a straightforward modification when
the orientifold action is modified to produce $SO(2N+1)$ or
$Sp(2N)$. For convenience, we provide a unified derivation of the
matrix measures for all the cases $SO(2N),Sp(2N),SO(2N+1)$ in an
Appendix. As reviewed in section \ref{sec_witten}, the latter groups arise
when discrete $B_{NS\,NS}$ or $B_{RR}$ torsion is introduced in 
spacetime. We will return to this in section \ref{sec_torsion}.

\section{Properties of Orientifolded Fermi Fluids}

\subsection{General Properties}

We now discuss some qualitative properties of the orientifolded fermi
fluid and associated $\half$-BPS geometries, using the standard
orientifold projection of $SO(2N)$ type. As we have seen, the allowed
fluid profiles have to be invariant under a rotation by $\pi$ about
the origin of phase space, or equivalently under $(x,p)\to
(-x,-p)$. The resulting space under this quotient is $\IC/Z_2$ and
can be represented by the upper half-plane $x>0$ along with half of
the $x=0$ axis, since points $(0,p)$ are identified with
$(0,-p)$. Fermi fluid profiles on this space can be represented either
as bubbles in a fundamental region, such as the upper half-plane, or as
$Z_2$-invariant configurations in the whole plane.

\DOUBLEFIGURE{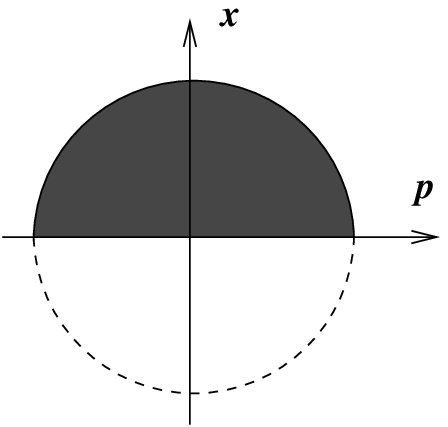}{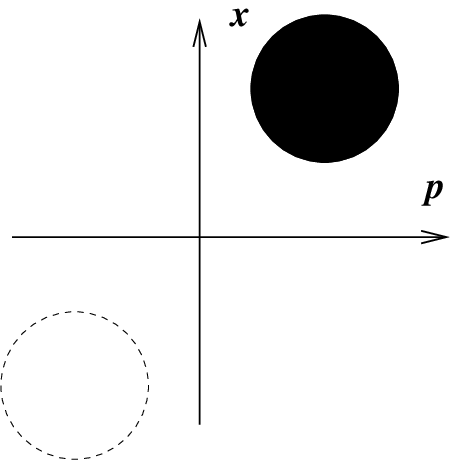}{Profile describing 
$AdS_5\times RP^5$.}{Profile describing $AdS_5\times S^5$, 
arising from D3 branes far from the orientifold.} 

The simplest example is the semi-circle centred at the origin, which
seeds the geometry $AdS_5\times RP^5$. As a fluid profile, it is the
ground state of $N$ free fermions in the semi-infinite harmonic
oscillator. We may note right away that the quotienting destroys
translational invariance in the $x-p$ plane and therefore the
semi-circle has to be centred at the origin (Fig. 1). 
Here and in what follows, the fundamental part of the bubble is shaded
while the $Z_2$ image is indicated with a dotted line to exhibit the
fact that in the ``upstairs'' space this is a $Z_2$ invariant
configuration.

We can also have circular configurations such as the one in Fig.2,
where no part of the circle touches the horizontal axis. This arises
as the near-horizon geometry of D3-branes parallel to, but far away
from, an orientifold plane. This example shows that geometries which
occur in the un-orientifolded case can also occur in the presence of
the orientifold. All that is required is for the boundaries to be
completely contained in the upper half-plane. The lack of
translational invariance, alluded to above, means that the distance of
the disc in this figure from the horizontal axis is physically
meaningful, in contrast to the un-orientifolded case.

\DOUBLEFIGURE{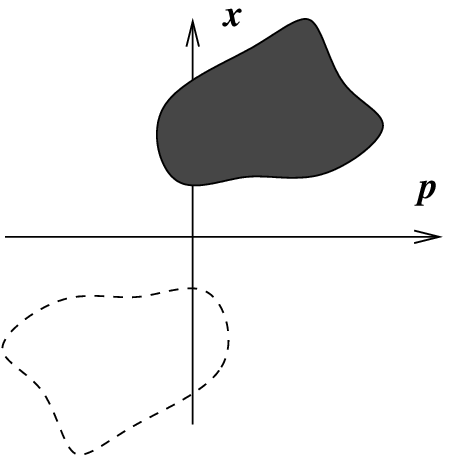}{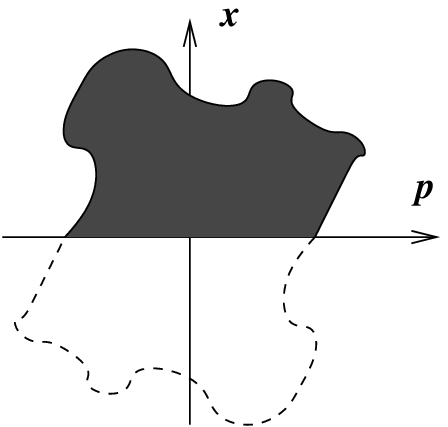}{Bubble excluding the origin.}{Bubble including the origin.}

More general configurations are the union of bubbles of two basic
types: those where the origin is included in the black region, and
those where it is not. Examples of the two types are shown in
Figs.3,4.  From Fig.4 we notice that a boundary component of the
latter type will be smooth only if, at the two points where it touches
the real axis, the slope is the same.

As noted earlier, when the origin is in a black region there is an
$RP^3$ in the spacetime geometry. This is the case for the bubble in
Fig.4. For bubbles that do not include the origin, at first sight
there appear to be two types: one illustrated in Fig.3 where the
bubbles are completely contained in the upper half plane (along with
their images in the lower half plane), and another type as in Fig.5
where the boundaries cross the horizontal axis at $x=0$. 
\FIGURE{
\epsfbox{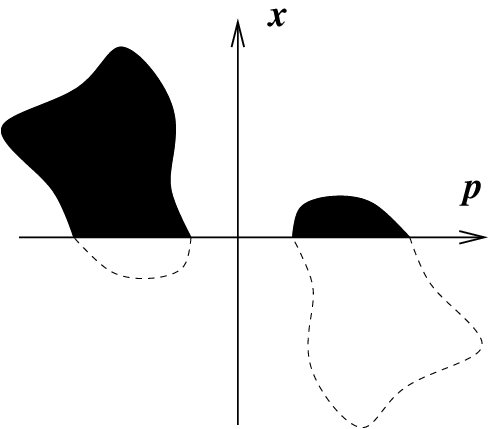}
\caption{Profile similar to Fig.3 after rotation.}}
The difference between these two examples is only superficial and can
be eliminated by making a different choice of fundamental
region. Choosing the right half plane as the fundamental region in
Fig.5 puts the bubble completely inside the region, with its image on
the other side. Thus we see that there are only two types of bubbles.
Deforming these two types into each other leads to topology change via
singular geometries, as we will see below.

In the models of Ref.\cite{Lin:2004nb}, particular types of bubbles
were identified with ``giant gravitons''. These are bubbles consisting
of a black region with a small hole inside, the area of the hole being
much smaller than the area of the black region surrounding it. The
corresponding geometries were interpreted as containing giant
gravitons made up of D3-branes wrapping a maximal $\tS^3$ in $S^5$. In
the orientifolded case we may instead consider a hole inside the bubble
of Fig.1. We see that there are two types of such holes,
and correspondingly two types of giant gravitons (Fig.6). If the small hole
is in a generic location then we have giant gravitons wrapping a 3-sphere
$\tS^3$ in $RP^5$. On the other hand if the hole surrounds the origin,
we have giant gravitons wrapping an $RP^3$ cycle of
$RP^5$. 
\FIGURE{
\epsfbox{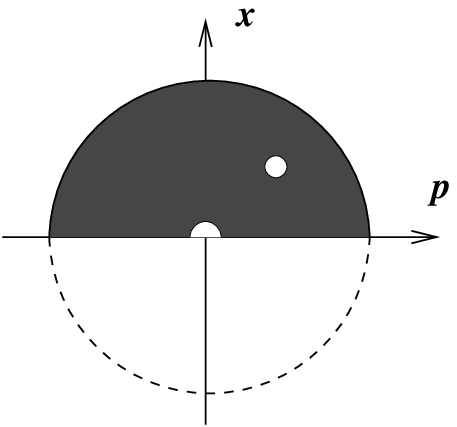}
\caption{Two types of giant gravitons in $AdS_5\times RP^5$.}}

Interestingly, as the hole around the origin becomes large enough to
interpret this as a new back-reacted geometry, we find that the $RP^3$
cycle has disappeared -- for the reason, stated earlier, that the
black region does not enclose the origin. Also, when the hole expands
further so that the black region forms a thin semicircular ring (stuck
to the horizontal axis), we can interpret the configuration as
consisting of dual giant gravitons wrapping an $S^3$ of $AdS_5$, and
uniformly distributed around an equator of $RP^5$.

\subsection{Torsion Cycles}\label{sec_torsion}

A unique feature of the $AdS_5 \times RP^5$ theory as compared to the
unorientifolded case is the appearance of {\it discrete torsion}, as
explained by Witten\cite{Witten:1998xy} and discussed in section
\ref{sec_witten}.  This means that there are topologically distinct
configurations of the $B$-field, described by the theta angles
$\theta_{NS\,NS}$ and $\theta_{RR}$.  In this section, we wish to
understand how this feature manifests itself in terms of the free
fermion description on the distinguished $(x_1,x_2)$ geometry-seeding
plane.

\EPSFIGURE{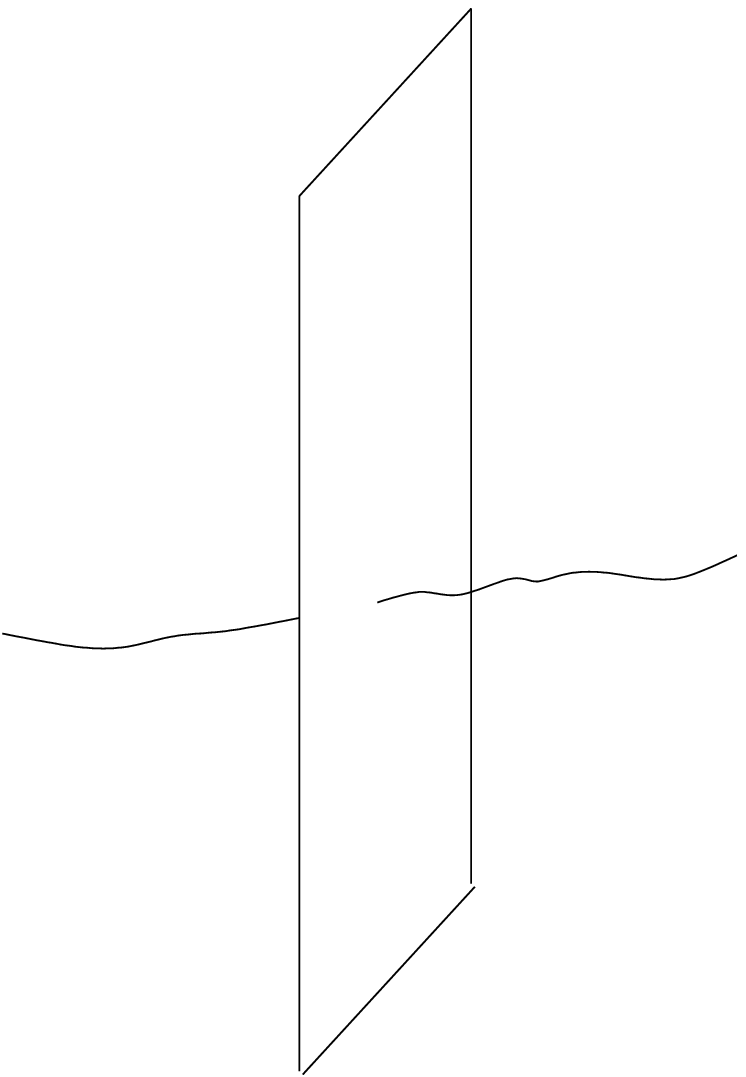,height=5cm}{The path $T$, with endpoints $P$ and $Q$,
                             intersects the $D3$ brane.}
As a warm-up,
consider first a $D3$ brane in $AdS_5 \times S^5$.
We may take it to be localized in the radial direction $\rho$ on $AdS_5$,
as well as located at a point on the $S^5$. It then extends along an
$S^3$ in the $AdS_5$. Suppose now that we choose a path $T$ whose endpoints
are two points $P$ and $Q$
on opposite sides of the 3-brane, as shown in Fig.7.
Consider the integral of $dF_5$ over the domain $T \times S^5$.
Since the brane provides a source for the five-form flux $F_5$,
and upon using Stokes's theorem, we find
\bea
  2\pi = \int_{T \times S^5} dF_5
       = \int_{P \times S^5} F_5 - \int_{Q \times S^5} F_5.
\eea
We can conclude that the flux through the $S^5$ changes by one
upon crossing the brane, changing the the gauge group from $SU(N)$
to $SU(N \pm 1)$. On the orientifold, the gauge group instead changes
from $SO(2N)$, $SO(2N+1)$ or $Sp(2N)$
to $SO(2N \pm 2)$, $SO(2N +1 \pm 2)$ or $Sp(2N \pm 2)$ respectively,
since $RP^5$ is covered twice by $S^5$.

>From the outset, the orientifolded theory has no preferred set of discrete
torsion, i.e. theta angles
$(\theta_{NS\,NS},\theta_{RR})$. However, once a particular set of values
has been assigned to some region, discrete torsions will automatically
be induced on the entire space, depending on the distribution of magnetic
sources. In the absence of magnetic sources
for the $B$-fields, the torsions are constant over the entire space.

However, $D5$ or $NS5$ branes wrapped on $RP^2$ cycles,
forming effective $D3$ branes on $AdS_5$, do provide such
sources. Crossing such a $D3$ brane will, in addition to changing the
rank of the gauge group as we just described, also cause the discrete
torsion to change.

More concretely, suppose we are wrapping a $D5$ brane on an $RP^2$ cycle\footnote{
The case of wrapping an $NS5$ brane on an $RP^2$ cycle is handled similarly,
and will lead to a domain wall across which $\theta_{NS \,NS}$ jumps.}.
This will result in an effective $D3$ brane which additionally has an
RR-form flux $H_{RR}=dB_{RR}$, which we choose to vanish to
preserve supersymmetry.  On the $AdS_5$ space, suppose the brane is
again localized in the radial direction, but extended in the $S^3$
directions. On the $RP^5$, the brane is extended along an $RP^2$. This
means that it must be localised on an $RP^3$, at least locally. The
$RP^2$ and the $RP^3$ are generically intersecting at one
point. Consider, then, the integral of $dH$ over $T \times RP^3$,
where $T$ is the same path as before.  Similar to the $dF_5$ integral
dealt with previously, we now find
\bea
  2\pi = \int_{T \times RP^3} dH
       = \int_{P \times RP^3} H - \int_{Q \times RP^3} H.
\eea
Hence, the theta angle $\theta_{RR}$ jumps upon crossing the brane.

Our objective is now to understand what this picture corresponds to
on the LLM plane $(x_1, x_2)$ of \cite{Lin:2004nb}.
An $AdS_5 \times RP^5$ background geometry
implies that we should start out with a ``semicircular'' black region
centred at the origin. The quotation marks indicate that the
correct designation of the geometry in the LLM plane requires
taking the non-trivial identifications into account.

In general, an $RP^i$ submanifold is realized in
$RP^j$, $i<j$, by the inclusion
\bea
  (x_1, \ldots, x_{i+1}) \rightarrow (x_1, \ldots, x_{i+1},0,\ldots,0).
\eea
To find an $RP^2$ inside $RP^5$, we must therefore choose three
coordinates $\{\tilde{X_1},\tilde{X_2},\tilde{X_3}\}$
among the six embedding coordinates $\{ X_i \}$, defined by
Eq.(\ref{embed}). The remaining $X_i$'s
should be set to zero.

For the $S^3$
of $AdS_5$ to separate two well-defined regions on the boundary plane,
it should not shrink to zero size there.
The magnetic $D3$ branes will then be associated with the white
region.
In this region, the radius
of the $\tilde{S}^3$ embedded in $S^5$ shrinks to zero,
corresponding to
$\sin \theta = 0$. Hence, $X_1 = X_2 = X_3 = X_4=0$. We must therefore
include at least one of the LLM coordinates in the generically non-vanishing
set $\{ \tilde{X}_i\}$, since otherwise the $X_i$'s cannot square
to unity. Actually, $X_5$ and $X_6$ are only proportional to the
LLM coordinates $x_1$ and $x_2$, differing by a $\rho$-dependent factor.
Recalling Eq.(\ref{polardefs}), we can write
\bea
  x_1 &= \sqrt{r_0}\, X_5 \cosh \rho \cr
  x_2 &= \sqrt{r_0}\, X_6 \cosh \rho.
  \label{eq_LLMrel}
\eea
This confirms that the brane will always be positioned outside of the
black ``semidisc'' of radius $r_0 \equiv R^2$.

Suppose first that precisely one of the two LLM coordinates belong
to the generically non-vanishing set. All the other embedding coordinates
must then vanish. Consequently, the non-vanishing coordinate, which
is either $X_5$ or $X_6$, must be assigned the value $R$.
Due to (\ref{eq_LLMrel}), this means that the brane appears as a point
at radial position $r_0 \cosh \rho$.

The other type of brane appears if we choose both of the LLM coordinates
to belong to the generically non-vanishing set. Again, all other embedding
coordinates vanish. The embedding condition then implies that
$X_5^2 + X_6^2=R^2=r_0$. Taking (\ref{eq_LLMrel}) into account, this leads us
to conclude that in this case, the brane appears as a ``semicircle''
of radius $r_0 \cosh \rho$. More precisely, the $RP^2$ collapses to
an $RP^1$ in the LLM plane.

\EPSFIGURE{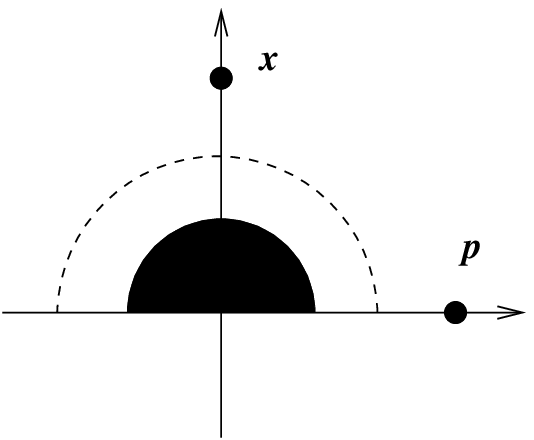,height=5cm}{Magnetically charged $D3$ branes
                             in an $AdS_5 \times RP^5$ background.}
In conclusion, we have found two different types of magnetic $D3$
branes, appearing on the LLM plane as illustrated in Fig.8.  One type
appears as points, restricted to lie on one of the axes.  The
positions of two such branes are indicated by black dots in the
figure.  The other type appears as $RP^1$ cycles, one of which is
shown.  Every such cycle divides the LLM plane into two parts,
allowing these D-branes to separate regions of unequal discrete
torsion $(\theta_{NS \, NS},\theta_{RR})$.  In both cases, the $RP^2$
is completely collapsed onto the LLM plane. In addition, the branes
extend along the $S^3$ directions of the $AdS_5$.

We have illustrated the ideas involved assuming
an $AdS_5 \times RP^5$
background, but our conclusions should be equally valid also for
the other \hBPS\ geometries. In particular, the topology on the LLM
plane, including the distribution of domain walls, is
expected to reflect the topology of the full bulk geometry.

\subsection{Topology Change}

In the absence  of orientifolds, topology change in the family of
\hBPS\ geometries of LLM type takes place at a fermi fluid
boundary satisfying a local equation\cite{Horava:2005pv} like
$x_1\,x_2=0$. This can be thought of as a limit of the smooth boundary
$x_1\,x_2=\mu$. Suppose this bounds a fermi fluid profile where for
$\mu>0$ there is a single black region extending to infinity in, say,
the upper left and lower right quadrants. Then for $\mu<0$ we have
instead two separated black regions, one extending to infinity on the
upper left and the other on the lower right. Clearly the connected
black region splits into two disconnected ones as we pass through
$\mu=0$ in the parameter space. It has been argued in
Ref.\cite{Horava:2005pv} that this is the ``irreducible'' transition
responsible for topology change, and that more general topology
changes can be decomposed into a sequence of such transitions.
The 10-dimensional geometry changes by the appearance or disappearance
of $S^5$ factors.
\DOUBLEFIGURE{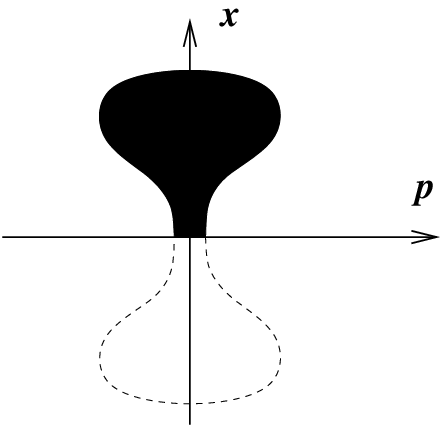}{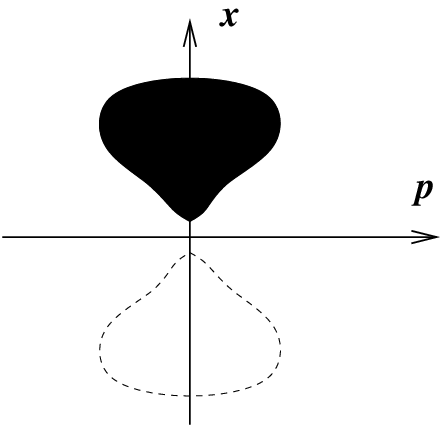}{Configuration of the topology of 
$AdS_5\times RP^5$, developing a neck.}{Configuration of the topology
of $AdS_5\times S^5$.} 

In the orientifolded theory there is another basic process of topology
change that creates or destroys cycles of order two. The basic process
of this type is the conversion of $RP^5$ into $S^5$ and vice
versa. What happens is that an $AdS_5\times RP^5$ configuration
consisting of a semi-disc anchored on the $x_1$ axis can be deformed
until it develops a narrow neck connecting it to its image, as shown
in Fig.9. At some stage the neck pinches off (Fig.10), and the bubble
is no longer anchored to the horizontal axis.
\FIGURE{
\epsfbox{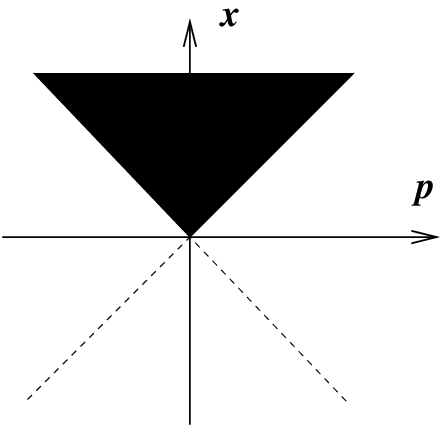}
\caption{Basic configuration for order-2 topology change.}}
Therefore this process represents a transition between $RP^5$ and
$S^5$. We see that the origin was contained inside the bubble in the
initial configuration, but is no longer contained at the end of the
process. As we saw earlier, the former type of configuration has an
$RP^3$ in the geometry while the latter does not. Thus this type of
topology change causes us to lose or gain cycles of order two. At the
transition, the (singular) configuration looks like a filled quadrant
along with its mirror image, as shown in Fig.11. We expect that there
will be topology-changing processes of ``order $n$'' built out of this
basic process, as in Ref.\cite{Horava:2005pv}.

\section{M-theory Orientifolds}

M-theory contains orientifold
5-planes\cite{Dasgupta:1995zm,Witten:1995em}, so one can try to apply
the construction above to this case. These orientifolds have been
investigated further in Refs.\cite{Hori:1998iv,Gimon:1998be}. In
particular one can take $N$ M5-branes parallel to an orientifold
5-plane, and the near-horizon geometry is $AdS_7\times
RP^4$\cite{Ahn:1998qe}. Unlike the D3-O3 system in type IIB string
theory, here the sphere factor becomes non-orientable after
quotienting. This is because the space transverse to the original
branes (and the orientifold plane) was $R^5$, whose orientation is
reversed upon reflection.

What is the effect of orientifolding on the $x_1 - x_2$ plane of the LLM
geometry? We cannot be as explicit as in the type IIB case. There we
had a precise map between this plane and the phase space of free
fermions. The orientifold plane was realised on the phase space as a
wall, blocking off the region of $x<0$ (or more precisely identifying
the lower half plane with a rotated copy of the upper half
plane). Therefore this was also the case in the $x_1 - x_2$ plane and
one could easily characterise functions $z$ which gave rise to the
general orientifolded \hBPS\ geometry -- as we did in the previous
sections. 

In the M-theory case, the free fermions are related to
\hBPS\ excitations of the $(2,0)$ theory on M5-branes and arise from
the transverse coordinates of the brane which are realised as scalar
fields of the $(2,0)$ theory\footnote{For a discussion of geometries
in a pp-wave background, see Ref.\cite{Bak:2005ef}.}. Since all the
transverse directions are being orientifolded, we again expect the
fermions to move in a semi-infinite harmonic oscillator. Though we
have seen in section \ref{briefreview} that the map from the fermion
phase space to the $x_1 - x_2$ plane is not a simple identification,
it was noted there that the topology of the two should be the
same. Therefore we expect that the orientifold plane in the M-theory
case is realised as the $x_1$ axis in the $x_1 - x_2$ plane of the
geometry\footnote{More precisely, there should be a choice of
conformal transformation which maps the orientifold to the $x_1$
axis. Thereafter we will only be allowed to make conformal
transformations that preserve the real axis, namely those which lie in
$SL(2,R)\subset SL(2,C)$.} just as for type IIB strings. In that case
most of the considerations in this paper will go through and we can
generate a precise collection of
\hBPS\ bubbling orientifolds in M-theory. We leave a more detailed 
analysis of this system for future work.

\section{Conclusions}

We have found an infinite class of new \hBPS\ geometries in type IIB
string theory and, somewhat less explicitly, in M-theory. These have
the same local geometry as the so-called ``bubbling geometries'' of
Ref.\cite{Lin:2004nb}, but have global identifications. The geometries
themselves have no orientifold plane and therefore the underlying
string theory has no open-string sector, but the identifications can
nevertheless be thought of as due to orientifold planes. The
geometries have cycles of order 2, some of which support discrete
torsion, in addition to the usual homology cycles. These backgrounds
have the same amount of supersymmetry and, upto discrete factors, the same
$SO(4)\times SO(4)$ symmetry as the original bubbling geometries. We
saw that they are associated to free fermions in a half-oscillator
potential, which in turn arise as the eigenvalues of matrices in the
Lie algebra of $SO(2N), Sp(2N)$ and $SO(2N+1)$.

It would be interesting to extend these results to the $\frac14$-BPS
geometries associated to $D3-D7-O3-O7$ systems. The introduction of
$D7$-branes leads to a varying axion-dilaton background and the
associated bubbling geometries are fairly simple and have already been
found in Ref.\cite{Liu:2004ru}. The introduction of orientifold
7-planes makes the system more interesting as it can then be related
to ``F-theory''
compactifications\cite{Vafa:1996xn,Morrison:1996na,Morrison:1996pp}.
It has been argued\cite{Sen:1996vd} that orientifold 7-planes are
dynamical and behave in some regions of moduli space like
non-perturbative D-branes (which at strong coupling exhibit remarkable
effects including, for example, exceptional global
symmetries\cite{Dasgupta:1996ij,Gaberdiel:1997ud}). In this context,
aspects of the AdS/CFT correspondence have been discussed in
Refs.\cite{Fayyazuddin:1998fb,Aharony:1998xz} and it should be
possible to find more general solutions of this kind using the
bubbling prescription.

Finally, as we mentioned at the beginning, orientifolded bubbling
geometries could be realised as local geometries of supersymmetric
flux compactifications. Whenever the fermi fluid is localised in a
bounded domain, the geometry is asymptotically $AdS_5\times RP^5$ (or
$AdS_5\times S^5$), and therefore can be matched on to flat
spacetime. But there might be a prescription as powerful as bubbling
(with its connection to free fermions) that describes genuine compact
geometries with orientifolds. This might make it much easier to
classify supersymmetric flux vacua.

\section*{Acknowledgements}

We are grateful to Oleg Lunin, Martin Olsson, Kyriakos Papadodimas and
Sandip Trivedi for helpful discussions.

\appendix
\section{Free fermions from matrix quantum mechanics}

In this appendix, we show that free fermions emerge from all the
traditional gauge groups. While some of this material is known or
implicit in the matrix model literature, it is useful to compile all
the needed results here.

The key observation is that
\bea\label{app_key}
  H &=& \sum_{i=1}^m{1\over\sqrt{g}}{\del\over\del \lambda_i}
  \sqrt{g}{\del\over\del \lambda_i}
  = \sum_{i=1}^m{1\over\Delta^2}{\del\over\del \lambda_i}
  \Delta^2{\del\over\del \lambda_i} \cr
  &=& {1\over \Delta}\,\sum_{i=1}^m{\del^2\over
  \del\lambda_i^2} \Delta,
\eea
where the measure factor $\sqrt{g}=\Delta^2$ is given by
\bea\label{app_measures}
  & (\Delta_-)^2  \hspace{2cm} & SU(m) \cr
  & (\Delta_-)^2(\Delta_+)^2(\Delta_0)^2  \hspace{2cm} & Sp(N), N=2m \cr
  & (\Delta_-)^2(\Delta_+)^2  \hspace{2cm} & SO(N), N=2m \cr
  & (\Delta_-)^2(\Delta_+)^2(\Delta_0)^2  \hspace{2cm} & SO(N), N=2m+1,
\eea
where we have defined 
\bea
  \Delta_- & \equiv \prod_{i<j}^m & (\lambda_i - \lambda_j) \cr
  \Delta_+ & \equiv \prod_{i<j}^m & (\lambda_i + \lambda_j) \cr
  \Delta_0 & \equiv \prod_{i}^m   & \lambda_i,
\eea
expressed in terms of the eigenvalues $\{ \lambda_i \}_{i=1,2,\cdots,m}$.
The last equality in (\ref{app_key}) follows from the identity
\bea
  \sum_{i=1}^m{\del^2\over
  \del\lambda_i^2} \Delta = 0.
\eea
The relation (\ref{app_key}) means that a factor of $\Delta$
can be absorbed by the wave function, giving rise to a system
of free fermions.

It remains to establish (\ref{app_measures}). Let us begin by
considering the unitary group, consisting of matrices
$U$ satisfying $UU^\dagger=\mathbb{I}$.
The Lie algebra $su(2)$ consists of anti-hermitian matrices
$A^\dagger=-A$. An infinitesimal variation $\delta U = U^{-1} dU$
of a unitary matrix $U$ is an element of the Lie algebra,
$\delta U^\dagger =- \delta U$.

The hermitian matrix A can be diagonalized 
using unitary matrices,
\bea\label{app_hdiag}
  U^{-1} A U = \Lambda = \mbox{diag}(\lambda_1, \lambda_2, \cdots,\lambda_m)
\eea
The significance is that we are trading degrees of freedom in
the anti-hermitian matrix for those of the eigenvalues themselves
and of the diagonalizing unitary transformation.
Differentiating equation (\ref{app_hdiag}) and solving for $dA$ gives
\begin{equation}\label{app_hmetric1}
    \mbox{Tr } (dA dA^\dagger) =
    \mbox{Tr } \left( 2 \delta U \Lambda \left[\delta U , \Lambda \right]
      + \left( d\Lambda \right)^2\right),
\end{equation}
which is manifestly invariant under unitary similarity transformations.
Hence it defines a metric on the space of antihermitian matrices.

Denoting matrix elements of a matrix $M$ by $M_{ij}$, this becomes
\begin{equation}\label{app_hmetric2}
    \mbox{Tr } (dA dA^\dagger) =
    2 \sum_{i<j}^m \delta u_{ij} \delta u_{ij}^* (\lambda_i - \lambda_j)^2
    + \sum_{i=1}^m d\lambda_i^2,
\end{equation}
We can read off the measure
\bea
  \sqrt{g} = (\Delta_-)^2,
\eea
(up to some numerical constant) in accordance with (\ref{app_measures}).

Next, we turn our attention to the symplectic gauge group
$\mbox{Sp}(2m)$, consisting of matrices $S$ satisfying
$S^T J S = J$, where
\bea\label{app_sLambda}
  J &=& \pmatrix{ 0 & \mathbb{I} \cr
                 -\mathbb{I} & 0},
\eea
where $\mathbb{I}$ is the $m \times m$ identity matrix.
The Lie algebra $\mbox{sp(2m)}$ then consists of matrices $A$ such that
$A^T=JAJ$.
Infinitesimal variations
$\delta S = S^{-1}dS$ belong to the Lie algebra $\mbox{sp}(2m)$.

Again, a Lie algebra element can be diagonalized using symplectic matrices,
$S^{-1}AS= \Lambda$.
Differentiating $S^{-1}AS= \Lambda$ and solving for $dA$ gives
\begin{equation}\label{app_smetric1}
    \mbox{Tr } (dA J dA^T J) =
    \mbox{Tr } \left( 2 \delta S \Lambda \left[\delta S , \Lambda \right]
      + \left( d\Lambda \right)^2\right).
\end{equation}

Defining $(e_{PQ})_{LM} = \delta_{PL} \delta_{QM}$,
a basis for $\mbox{sp}(2m)$ is
\bea
  e_{j,k}^1 & \equiv e_{j,k} - e_{k+m,j+m} \cr
  e_{j,k}^2 & \equiv e_{j,k+m} + e_{k,j+m} \cr
  e_{j,k}^3 & \equiv e_{j,k} - e_{k+m,j+m}.
\eea
Note that since
$0 = \det (A-\lambda \mathbb{I}) = -\det(A+\lambda \mathbb{I}) $,
eigenvalues come in pairs, $\{ (\lambda_i,-\lambda_i)  \}_{i=1,2,\cdots,m}$.
Consequently, we can write $\Lambda$ on the form 
\bea\label{app_slambda}
  \Lambda = \sum_{i=1}^k \Lambda_i e_{ii}^1 =
  \mbox{diag} (\lambda_1, \lambda_2, \cdots, \lambda_m, -\lambda_1, -\lambda_2, \cdots, -\lambda_m).
\eea
Similarly, we write $\delta S$ as
\bea
  \delta S & = & \sum_{i=1}^m \delta S_i^d e_{ii}^1 \cr
  & + & \sum_{j<k}^m \left[ \delta S_{jk}^i (e_{jk}^1 + e_{kj}^1)
                      +\delta S_{jk}^{ii} (e_{jk}^1 - e_{kj}^1)  \right] \cr
  & + & \sum_{j \le k}^m \left[ \delta S_{jk}^a (e_{jk^2 + e_{jk}^3})
                      +\delta S_{jk}^b (e_{jk}^2 - e_{jk}^3)  \right].
\eea
The reason for this particular expansion is that it will make the metric
diagonal. Indeed, differentiating $S^{-1}AS= \Lambda$ and
solving for $dA$ gives
\bea
  \mbox{Tr } (dA^T J dA J) & =
  2 & \sum_{i=1}^m d\lambda_i^2 \cr
 & +8 & \sum_{j < k} 
    (\lambda_j - \lambda_k)^2 \left[ (\delta S_{jk}^{ii})^2 - (\delta S_{jk}^{i})^2 \right] \cr
 & +8 & \sum_{j \le k}
   +(\lambda_j + \lambda_k)^2 \left[ (\delta S_{jk}^{b})^2 - (\delta S_{jk}^{a})^2 \right].
\eea
Hence, the measure is
\bea
  \sqrt{g} = (\Delta_-)^2(\Delta_+)^2(\Delta_0)^2,
\eea
as anticipated in (\ref{app_measures}).

The orthogonal gauge group remains to be analysed.
Orthogonal matrices $O$ satisfy
$OO^T = \mathbb{I}$, and corresponding Lie algebra elements are
antisymmetric, $A A^T=-A$.
The infinitesimal variation $\delta O = O^{-1} dO$ of an orthogonal
matrix $O$ is antisymmetric, $\delta O^T = - \delta O$.

As in the symplectic case, the eigenvalues come in pairs, differing only
by a sign. This means that $SO(2m)$ and $SO(2m+1)$ both have $m$
independent eigenvalues. Consider first the case $SO(2m)$.
Using orthogonal similarity
transformations, antisymmetric matrices can only be brought to the
block-diagonal form
\bea\label{app_oLambda}
\Lambda &=& \pmatrix{ \lambda_1\pmatrix{0&1\cr -1 &0} & 0 & \cdots &0\cr
0 & \lambda_2\pmatrix{0&1\cr -1 &0} & \cdots &0\cr
\cdots &\cdots &\phantom{\pmatrix{0&1\cr -1 &0}}&\cdots\cr
0 & 0 & \cdots & \lambda_N\pmatrix{0&1\cr -1 &0}}\cr
&=& i\sigma_2\otimes \diag(\lambda_1,\lambda_2,\ldots,\lambda_N).
\eea

Suppose that $O$ is the orthogonal matrix which block-diagonalizes
the antisymmetric matrix $A$, i.e.
\begin{equation}\label{app_blockdiag}
  O^{-1}AO = \Lambda.
\end{equation}
Differentiating equation (\ref{app_blockdiag}) and solving for $dA$ gives
\begin{equation}\label{app_ometric1}
    \mbox{Tr } (dA dA^T) =
  - \mbox{Tr } \left( 2 \delta O \Lambda \left[\delta O , \Lambda \right]
      + \left( d\Lambda \right)^2\right).
\end{equation}

Defining
\bea
  \sigma^0 & \equiv \mathbb{I}_{2 \times 2}  = \pmatrix{+1&0\cr 0 &+1} \hspace{1.5cm}
  \sigma^2 & \equiv i\sigma_2  = \pmatrix{0&+1\cr -1 &0}      \cr
  \sigma^1 & \equiv \sigma_1  = \pmatrix{0&+1\cr +1 &0}       \hspace{1.8cm}       
  \sigma^3 & \equiv \sigma_3  = \pmatrix{+1&0\cr 0 &-1}
\eea
as a basis of real $2 \times 2$ matrices in terms of the identity matrix
$\mathbb{I}_{2 \times 2}$ and the Pauli matrices $\{ \sigma_i \}$, we can write
the variation $\delta O$ as
\bea\label{app_deltaO}
\delta O &=& \pmatrix{ \delta O_{11}\pmatrix{0&1\cr -1 &0} & \delta
O_{12}^A 
\sigma^A & \cdots & \delta O_{1N}^A \sigma^A\cdots \cr
-\delta O_{12}^A \sigma^A & \delta O_{22}\pmatrix{0&1\cr -1 &0} & 
\cdots &\delta O_{2N}^A \sigma^A\cr
\cdots &\cdots &\phantom{\pmatrix{0&1\cr -1 &0}}&\cdots\cr
-\delta O_{1N}^A \sigma^A & -\delta O_{2N}^A \sigma^A & \cdots & 
\delta O_{NN}\pmatrix{0&1\cr -1 &0}} \cr
&=&   \sigma^A \otimes \delta O^A
    + i\sigma_2\otimes \diag(\delta O_{11},\delta O_{22},\ldots,\delta O_{NN}).
\eea
The antisymmetric matrix $\delta O^A$ is defined such that the entry at
row $i$ and column $j$ is the coefficient $\delta O_{ij}^A$.
We are using the convention that repeated indices $A$ are summed over.

Using the forms (\ref{app_oLambda}) and (\ref{app_deltaO}),
the metric (\ref{app_ometric1}) becomes
\bea\label{app_ometric2}
\mbox{Tr } (dA dA^T) =
&-&4 \sum_{i<j} \sum_{A=1,3} \left( \delta O_{ij}^A \right)^2 
(\lambda_i+\lambda_j)^2 \cr
&+&4 \sum_{i<j} \sum_{A=0,2} (-1)^{A/2+1} \left( \delta O_{ij}^A \right)^2 
(\lambda_i-\lambda_j)^2
+ \sum_{i=1}^m d \lambda_i^2 
\eea
Consequently, the measure is
\bea\label{app_omeasure}
  \sqrt{g} = (\Delta_-)^2(\Delta_+)^2,
\eea
as written in (\ref{app_measures}).

Consider now the gauge group $SO(2m+1)$.
The analysis goes through in much the same way as for the $SO(2m)$ case,
replacing
\bea
  \delta O \rightarrow \delta O + \sum_{i=1}^{m} \sum_{B=0,1} \delta O_i^B \tau^B \otimes e_{i,2k+1},
\eea
where $\tau^0 \equiv \pmatrix{ 1 \cr 0}$,
$\tau^1 \equiv \pmatrix{ 0 \cr 1 }$.
Similarly, $\Lambda$ is just extended with another row and column of zeros,
since there are still only $m$ eigenvalues.
With these modifications, the metric (\ref{app_ometric2}) gets an
additional contribution,
\bea
  \mbox{Tr } (dA dA^T) \rightarrow \mbox{Tr } (dA dA^T)
    + \sum_{i,B} \lambda_i^2 (\delta O_i^B)^2.
\eea
The measure (\ref{app_omeasure}) changes accordingly,
\bea
  \sqrt{g} \rightarrow \sqrt{g} \prod_{i=1}^m \lambda_i^2 = 
  (\Delta_-)^2(\Delta_+)^2 (\Delta_0)^2,
\eea
consistent with (\ref{app_measures}).

Note that the symplectic and orthogonal gauge groups
share the following two features.
The independent
set of eigenvalues can always be chosen to be non-negative.
In addition, absorbing a factor of $\Delta$ into the wave function
makes the eigenvalues $\lambda_i$ repel not only each other,
but also their images $-\lambda_i$.
This is
consistent with the orientifold interpretation, as explained in
section \ref{sec_fermi}.

\end{document}